\documentclass[aps,prb,twocolumn,showpacs,superscriptaddress,floatfix,10pt,longbibliography,bibnotes,footinbib]{revtex4-2} 
\usepackage[utf8]{inputenc}
\usepackage[T1]{fontenc}
\usepackage{amsmath}
\usepackage{booktabs} 
\usepackage[caption = false]{subfig}
\usepackage{color}
\usepackage[dvipsnames]{xcolor}
\usepackage{braket}
\usepackage{bm}
\usepackage{dsfont}
\usepackage{adjustbox}
\usepackage{graphicx}
\usepackage{enumitem}
\usepackage{bbm}
\usepackage{tikz}
\usepackage{physics}
\usepackage{microtype}

\usepackage{txfonts}

\usepackage[unicode=true,colorlinks=true,pdfpagemode=UseOutlines,pdfstartview=Fith]{hyperref}
\hypersetup{linkcolor=blue,citecolor=blue,urlcolor=blue}

\usepackage{slashed}
\usepackage{orcidlink}

\makeatletter

\renewcommand{\Im}{\operatorname{Im}} 

\newcommand{\hc}{\mathrm{h.c.}} 

\newcommand{\px}{\partial_x}
\newcommand{\py}{\partial_y}
\newcommand{\enm}{\epsilon_{nm}}
\renewcommand{\bm}{\mathbf}
\renewcommand{\Im}{\text{Im}}

\newcommand{\bvec}[1]{{\bm{#1}}}

\usepackage[normalem]{ulem}

\makeatother

\begin{document}

\title{Raman circular dichroism and  spinon  quantum geometry of chiral quantum spin liquid}

\author{Eduard Koller \orcidlink{0000-0002-4939-7470} }
\thanks{eduard.koller@tum.de}
\affiliation{Technical University of Munich, TUM School of Natural Sciences, Physics Department, 85748 Garching, Germany}
\affiliation{Munich Center for Quantum Science and Technology (MCQST), Schellingstr. 4, 80799 M{\"u}nchen, Germany}
\affiliation{Institute for Advanced Study, Technical University of Munich, 
Lichtenbergstr. 2a, 85748 Garching, Germany}

\author{Valentin Leeb \orcidlink{0000-0002-7099-0682}}
\affiliation{Technical University of Munich, TUM School of Natural Sciences, Physics Department, 85748 Garching, Germany}
\affiliation{Munich Center for Quantum Science and Technology (MCQST), Schellingstr. 4, 80799 M{\"u}nchen, Germany}
\affiliation{Department of Physics, University of Zürich, Winterthurerstrasse 190, 8057 Zürich, Switzerland}

\author{Natalia B. Perkins}
\affiliation{School of Physics and Astronomy, University of Minnesota, Minneapolis, MN 55455, USA}
\affiliation{Institute for Advanced Study, Technical University of Munich, 
Lichtenbergstr. 2a, 85748 Garching, Germany}

\author{Johannes Knolle}
\affiliation{Technical University of Munich, TUM School of Natural Sciences, Physics Department, 85748 Garching, Germany}
\affiliation{Munich Center for Quantum Science and Technology (MCQST), Schellingstr. 4, 80799 M{\"u}nchen, Germany}
\affiliation{Blackett Laboratory, Imperial College London, London SW7 2AZ, United Kingdom}

\date{October 5, 2026}
\begin{abstract}
We show that the  quantum geometry of fractionalized spin excitations in Mott-insulating quantum spin liquids (QSL) gives rise to  a finite Raman circular dichroism (RCD) signal.  
 We identify conditions under which the Loudon--Fleury framework,  where light modulates the underlying exchange interactions, and the light-matter-coupling approach,  where the effective vector potential couples directly to the spinon bands,
yield equivalent Raman vertices for the fractionalized quasiparticles.  
Using the latter, we derive an analytical decomposition of the RCD into different quantum geometric contributions. This reveals the sensitivity of the RCD to the underlying structure of the wave functions and the handedness of the excitations, rather than a nonzero Chern number of spinon excitations.
To illustrate this, we apply our approach to two examples, the Kitaev honeycomb model in a magnetic field and a  parton mean-field state of the chiral QSL on the triangular lattice. 
\end{abstract}

\maketitle
\section{Introduction}
Topology and geometry are fundamental concepts in modern condensed matter physics governing the behavior and emergent properties of quantum materials \cite{Berry1984,Toermae2023,yu2024quantum}.
A key topological quantity is the Berry curvature, which emerges from the geometric phase acquired by the  wavefunction as a continuous parameter, for example time in adiabatic evolution from crystal momentum for Bloch states, changes~\cite{Berry1984,xiao2010berry}. It explains a wide range of phenomena, including
the quantum Hall effect \cite{Thouless1982,Niu1985}, topological insulators \cite{Kane2005,Sheng2006} and differences between integer and half-integer spin systems \cite{Berry1984,Canali2003}.
Berry curvature is manifest in various experimental observables, i.e.~governing transport such as the quantized Hall conductance in integer Hall effects~\cite{Klitzing1980,Laughlin1981,Thouless1982}, the thermal Hall response~
\cite{Kane1997,Kitaev2006,Kasahara2018}, as well as the orbital magnetic susceptibility~\cite{Raoux2015}.
However, the Berry curvature represents
only one aspect of a broader topological classification of wavefunctions. The quantum geometric tensor (QGT)  \cite{Provost1980,Berry1984}, also known as the Fubini--Study metric \cite{Study1905, Fubini1904}, provides a complete framework. 
The imaginary part of the QGT corresponds to the Berry curvature, while the real part corresponds to the quantum metric \cite{Piechon2016,yu2024quantum}, quantifying the distance between quantum states.
In recent years, the quantum metric has gained increasing attention as a tool to understand various  properties of quantum materials~\cite{Toermae2023,yu2024quantum,jiang2025revealingquantumgeometrynonlinear}, for example the localization of Wannier functions in band insulators
~\cite{Marzari1997,Marzari2007,Marzari2012} or contributions to the orbital magnetic susceptibility of Bloch bands, quantifying how electronic wavefunctions couple to external fields~\cite{Raoux2015,Piechon2016}. Furthermore, it is essential for understanding non-linear optical effects~\cite{gao2014field,morimoto2016topological,ahn2022riemannian} and superconductivity and superfluid weights in flat band systems~\cite{Peotta2015,Julku2016,Liang2017,Toermae2018,Arbeitman2022,Toermae2022,Peotta2023}. 

Research on QGT has focused on {\it electronic single-particle} wavefunctions with recent attempts to generalize it to many-body wavefunctions~\cite{salerno2023drude,wu2024corner}. However, many strongly interacting  many-body systems are described by  emergent low-energy quasiparticles, whose properties can be characterized by their QGT~\cite{fradkin2013field}. This raises the question, if and how the QGT manifests in experimental response functions.
Recently, it was shown that the optical response of topological magnon insulators~\cite{mcclarty2022topological} is a  direct way of probing the QGT of spin wave excitations~\cite{Bostroem2023}. Here, we show that even in quantum magnets with {\it emergent fractionalized excitations}, 
 the response to inelastic light scattering is governed by their quantum geometry.

QSL are long-range entangled phases of insulating quantum magnets without conventional magnetic order~\cite{savary2016quantum,Knolle2019}. They can be described in terms of parton theories of exotic excitations, which carry fractional quantum numbers of simple spin flips, interacting via dynamical gauge fields~\cite{fradkin2013field}. In many cases~\cite{wen2004quantum}, most prominently in the celebrated exactly solvable Kitaev honeycomb model~\cite{Kitaev2006}, QSL properties can be understood in terms of charge neutral weakly interacting single-particle excitations. 
For example, a time-reversal symmetry (TRS) breaking {\it chiral} QSL can be understood as
 a gapped phase hosting gapless chiral fermionic edge states, analogous to electronic quantum Hall systems~\cite{kalmeyer1987equivalence,fradkin2013field}, leading to a quantized thermal Hall response~\cite{Kitaev2006}. 
Among the available dynamical probes, Raman scattering has emerged as an effective tool for investigating QSLs,  providing direct access to energy-resolved signatures of fractionalized excitations, including Majorana fermions and more general spinon excitations~\cite{Knolle2014,Perreault2015,Nasu2016,Rousochatzakis2019PRB,Sandilands2015,Glamazda2016,Sahasrabudhe2020,Wulferding2020,ko2010raman}. The capability of Raman scattering to directly couple to spinons without exciting gauge fields~\cite{Knolle2016} makes it particularly valuable for studying their QGT.

\begin{figure}[t]
    \includegraphics[width=0.35\textwidth]{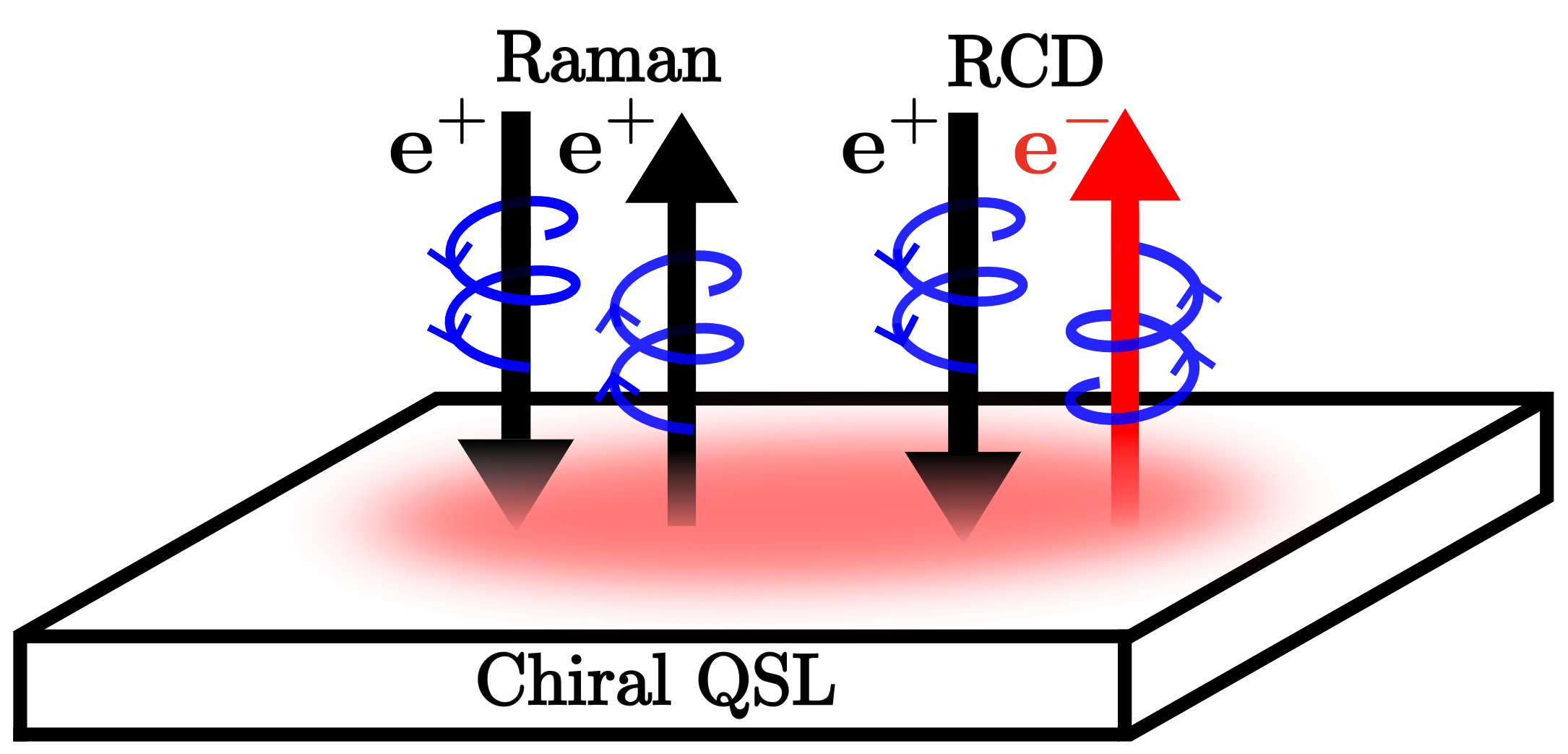}   
    \caption{\label{fig:sketch_RCD}
    Illustration of Raman Circular Dichroism (RCD) (right) and conventional Raman scattering (left).
    In RCD, circularly polarized light undergoes a polarization change during the scattering process by exciting chiral excitations. In contrast, conventional Raman scattering occurs without a change in polarization.
    }
\end{figure} 

In this  paper,  we demonstrate that Raman circular dichroism (RCD), defined as the difference in Raman scattering intensity between left- and right-circularly polarized light (Fig.~\ref{fig:sketch_RCD}), is  governed by   the  quantum geometry of spinons in  QSLs. 
We establish the connection by identifying conditions under which the Raman vertex obtained from the standard microscopic Loudon--Fleury approach~\cite{LF1968}, based on the light-induced modulation of exchange interactions, is equivalent to that derived from a direct light-matter coupling to the effective spinon bands~\cite{Topp2021}. 
We investigate the validity of this correspondence for two paradigmatic chiral QSL examples: the exactly solvable Kitaev honeycomb QSL in a field~\cite{Kitaev2006} and the chiral QSL for the antiferromagnetic Heisenberg model on the triangular lattice as described within a parton mean-field theory~\cite{kaneko2014gapless,Iqbal2016,Gong2019}.
We show that the RCD arises from the interplay of three distinct contributions: one proportional to the Berry curvature,  another to the quantum geometric connection, and a third involving 
higher-order momentum derivatives of the band projectors.
Consequently, the frequency dependence of the RCD reflects the interplay between the spinon density of states and the quantum geometry entering the Raman matrix elements.

\section{Raman response and light-matter coupling} 
The Loudon--Fleury (LF) formalism is a well-established exchange-scattering approach for calculating the Raman response  in Mott insulators~\cite{LF1968,Shastry1990,Shastry1991}. It relies on the similarity between the Raman response and the exchange interaction $J^{\alpha\beta}_{ij}$ for a general spin Hamiltonian, where virtual electron hopping is partially assisted by photons~\cite{LF1968,Shastry1990,Shastry1991,Devereaux2006,Fu2017,Yang2021}. The electronic degrees of freedom can be integrated out to obtain an effective coupling of spin degrees of freedom to light via the general Raman vertex
\begin{align}
    \label{eq:def_LF_R}
    R^{ss^\prime}_\text{LF} = \sum_{i,j}  ( \bvec{e}^{s} \cdot \bvec{d}_{ij}) ( \bvec{e}^{*s^\prime} \cdot \bvec{d}_{ij}) \sum_{\alpha,\beta} S^{\alpha}_i J^{\alpha\beta}_{ij} S^{\beta}_{j}. 
\end{align}
Here, $\bvec{e}^{s}$ and $\bvec{e}^{s^\prime}$  denote the polarization vectors of the  incoming and outgoing photons and $\bvec{d}_{ij}$ is the vector connecting spins at sites $i, j$.

Another approach to computing the Raman response,
commonly used for electronic tight binding models, is the light-matter coupling (LMC) formalism~\cite{Topp2021}. This framework directly couples light to matter excitations  and naturally connects the response to the band topology.
Within this phenomenological framework, minimal coupling  is applied to the effective single-particle Hamiltonian $h(\bvec{k}) \rightarrow h\left( \bvec{k} +e  \bvec{A}  \right)$, where $\bvec{A}$ is the vector potential. Thus, the two approaches start from different levels of description:
LF couples light to the microscopic exchange processes of the Mott
insulator, whereas LMC couples light directly to the effective
quasiparticle Hamiltonian.

The LMC tensors are extracted from an expansion of the Hamiltonian in the vector potential,
\begin{align}
    \label{eq:def_LMC-exp}  h\left( \bvec{k} +e \bvec{A}  \right) = h(\bvec{k}) + e \sum_\mu l_\mu^{(1)}(\bvec{k}) A_\mu + \frac{e^2}{2} \sum_{\mu,\nu } l_{\mu\nu}^{(2)}(\bvec{k})A_\mu A_\nu +...
\end{align}
The first-order term, 
$l_\mu^{(1)}$, corresponds to single-photon processes such as spontaneous emission or absorption, while the lowest-order contribution to Raman scattering arises from the two-photon process described by
$l_{\mu\nu}^{(2)}(\bvec{k})$, which defines the Raman operator 
\begin{align}
    \label{eq:def_LMC_R}
    R^{ss^\prime}_\text{LMC} = \sum_{\mu,\nu} \sum_{m,n}  e^{s}_\mu  e^{*s^\prime}_\nu \sum_\bvec{k}  \psi^\dagger_{\bvec{k},m} 
    l_{mn,\mu\nu}^{(2)}(\bvec{k})
    \psi_{\bvec{k},n}. 
\end{align}
Here, $l_{mn,\mu\nu}^{(2)}(\bvec{k}) = P_m(\bm k) \left[ \partial_{k_\mu} \partial_{k_\nu}h(\bvec{k})\right] P_n(\bm k)$  are the second-order LMC matrix elements,  with  $P_n(\bm k)  = |u_{n}(\bm k) \rangle \langle u_n(\bm k) |$ projecting onto the $n$-th band and $\psi_{\bm k}$  the fermionic fields.
In the case of weakly interacting electronic systems $\psi_{\bm k}$ correspond to the Bloch states. 
 By contrast, Ref.~\cite{Bostroem2023} applied the LMC formalism to charge-neutral magnons, where minimal coupling does not directly apply.

Using two example QSLs and a standard mapping of spins $S_i^\alpha$ to emergent low-energy fermionic spinon operators $\psi_{\bm k}$, we have derived the Raman vertex in both formulations: Eq.~\eqref{eq:def_LF_R}, the LF approach, starting from the Mott insulator and Eq.~\eqref{eq:def_LMC_R}, the LMC approach,  starting from the effective, charge-neutral, quasiparticle Hamiltonian, see App.~\ref{App:LFvsLMC}. 
The two approaches yield identical Raman vertices if the effective spinon kinetic energy $h(\mathbf k)$ is generated by the same exchange interactions that defines the LF Raman vertex Eq.~\eqref{eq:def_LF_R}. In this case,  
\begin{align}
R^{ss'}_{\rm LF} = -R^{ss'}_{\rm LMC}, \label{eq:LF_equiv_LMC}
\end{align} 
up to an overall sign which does not affect the Raman intensity. Additional contributions to the spinon kinetic energy originating from microscopic terms other than exchange interactions  generally enter the LMC vertex but are not captured by the standard exchange-based LF approach.  
A relevant example is the field-induced $\kappa$ term in the Kitaev model. 
While this term contributes to the effective Majorana hopping Hamiltonian and therefore enters the LMC description, its microscopic origin is the Zeeman coupling rather than an exchange process generated by virtual electron hopping.
Thus, the $\kappa$ term provides a simple example for which the LF--LMC correspondence does not hold.
We discuss this limitation of the LF--LMC correspondence in more detail in App.~\ref{App:LFvsLMC}.

For the exchange-generated component of the Raman vertex discussed above, the LMC formulation provides a transparent link between RCD and QGT of the underlying quasiparticle bands. In the following, we first recap details of the connection between LMC and RCD as well as the QGT for a minimal two band systems, as sufficient for our concrete examples.

\section{Raman Circular Dichroism and QGT}
The RCD quantifies the difference in the Raman intensities between left circularly $\bvec{e}^+$ and right circularly $\bvec{e}^-$ polarized light 
\begin{align}
    \label{eq:def_RCD}
    I_{\text{RCD}}(\omega) =  I^{+-}(\omega) - I^{-+}(\omega), 
\end{align} where the photon polarization states are given by 
$\bvec{e}^+= (1, i)^T/\sqrt{2}$
and  $\bvec{e}^- = (\bvec{e}^+)^*$. 
A finite RCD arises from TRS breaking, as the Raman intensities $I^{+-}(\omega)$ and $I^{-+}(\omega)$ are related by TRS.  It can  appear in conventional magnetically ordered systems. For example,  
in ferromagnets with spin anisotropies 
induced by
spin-orbit coupling, spin waves can have a preferred handedness \cite{onoda2008left,Jenni2022}, and the lack of an oppositely polarized counterpart naturally results in a non-zero  circular dichroism. In antiferromagnets, the two degenerate magnon branches have opposite handedness canceling the RCD, but in the presence of an external magnetic field or spin-anisotropic interactions, this degeneracy is lifted, resulting in a finite RCD~\cite{Hoffman2005,Bostroem2023}. RCD has also been discussed in the context of fractional quantum Hall (FQH) states where time-reversal and chiral symmetry are broken~\cite{liou2019chiral,nguyen2021probing}, with recent observations using circular polarized light~\cite{liang2024evidence}.
In the following, we show how the RCD is related to the QGT of spinons in chiral QSL.

First, we express the RCD in terms of LMC matrix elements.  
To this end, we consider a minimal two-band spinon Hamiltonian
\begin{equation}
H = \sum_{\bm k} \sum_{l=1,2} \epsilon_l(\bm k)\, P_l(\bm k)
= \sum_{\bm k} \epsilon_1(\bm k)\, P_1(\bm k) + \epsilon_2(\bm k)\, P_2(\bm k),
\end{equation}
where $\epsilon_l(\bm k)$ are the corresponding eigenenergies. We label the bands such that $\epsilon_1(\bm k)$ corresponds to the lower band and $\epsilon_2(\bm k)$ to the upper band.
By applying Eq.~\eqref{eq:def_RCD} along with the LMC Raman operator from Eq.~\eqref{eq:def_LMC_R}, we obtain
\begin{align}
    \label{eq:RCD_in_LMC}
    I_{\text{RCD}}(\omega) = &\sum_\bvec{k} 2 \text{Im} \left[ \left(l_{12,xx}^{(2)}(\bvec{k}) - l_{12,yy}^{(2)}(\bvec{k}) \right) l_{12,xy}^{*(2)}(\bvec{k})\right]  \nonumber \\
    &\times \delta\Big(\omega - 2 \epsilon(\bvec{k})  \Big),
\end{align}
where we focused on the particle-hole symmetric case $\epsilon(\bm k) \equiv  \epsilon_2(\bm k) = - \epsilon_1(\bm k)$ for the chiral QSLs considered, see App.~\ref{App:ExpansionRCD}. 

Having expressed the RCD in terms of LMC matrix elements, we now relate it to the band geometry. 
Since the LMC vertices involve momentum derivatives of the Hamiltonian, they can be recast 
in terms of derivatives of the band projectors, naturally connecting the response to geometric quantities.
We therefore introduce the quantum geometric tensor (QGT)~\cite{Berry1984,Provost1980}:
\begin{align}
\label{eq:QGT}
          Q^{mn}_{\mu \nu }(\bvec{k}) =\Tr \left[P_n(\bm k) \partial_\mu P_m(\bm k) \partial_{\nu} P_n(\bm k)  \right]& 
          = g^{mn}_{\mu \nu}(\bvec{k}) - \frac{i}{2} F^{mn}_{\mu \nu}(\bvec{k}),
\end{align}
where  $\partial_\mu \equiv \partial_{k_\mu}$.  The QGT encodes both the quantum metric $g_{\mu \nu}(\bvec{k})$ and the Berry curvature $F_{\mu \nu}(\bvec{k})$. 

In addition, higher-order derivatives give rise to the quantum geometric connection (QGC)~\cite{ahn2022riemannian,Mitscherling2025}
\begin{align}
\label{eq:QGC}
C^{mn}_{\eta; \mu \nu}(\bm k)  = 
\Tr\left[
P_n(\bm k) \partial_{\mu}P_m(\bm k) 
\left( \partial_{\eta} \partial_{\nu} P_n (\bm k)+ \partial_{\eta} P_m(\bm k) \partial_{\nu}P_n(\bm k)  \right) 
\right]. 
\end{align}
For a non-degenerate two-band system,  Eqs.~\eqref{eq:QGT} and \eqref{eq:QGC} reduce to their single-band forms~\cite{Mitscherling2025}.

Third, following Ref.~\cite{Topp2021}, the LMC matrix elements $l_{mn,\mu\nu}^{(2)}(\bvec{k})$ can be expressed in terms of derivatives acting on eigenstates, 
thereby relating them to the components of the QGT and QGC,  see App.~\ref{App:ExpansionRCD}. As a 
consequence, the RCD in Eq.~\eqref{eq:RCD_in_LMC} can be rewritten in terms of the Berry curvature, the imaginary part of the QGC, and additional higher-order derivative terms as
\begin{align}  
        &I_{\text{RCD}}(\omega) = I_\text{B}(\omega ) + I_\text{C}(\omega ) + I_\text{S}(\omega )  \label{eq:RCD_expansion_simple}   \\
        &=  \sum_\bvec{k}  \Big[
            v_\text{B}(\bvec{k} ) F_{xy}(\bvec{k})  +  v_\text{C}(\bvec{k} ) C(\bm k) 
             + v_\text{S}(\bvec{k} ) \Big] \delta \Big(\omega - 2 \epsilon(\bvec{k} )  \Big) \nonumber.
\end{align}
The RCD is related to the Berry curvature $F_{xy}(\bvec{k})$ and to a subset of multiple different components of the quantum connection $C^{mn}_{\eta; \mu \nu}(\bm k)$ via the  prefactors $v_\text{B}(\bvec{k})$ and $v_\text{C}(\bvec{k})$, which only depend on the energy or its derivatives.
The full analytical expression of the different components contributing and the prefactors are given in App.~\ref{App:ExpansionRCD}.  The third contribution $v_\text{S}(\bvec{k})$ is related to the second derivatives of the band projectors, such as $\partial_\mu \partial_\nu P_n(\bm k)$.

Before computing the RCD for specific examples, we first highlight two generic features.
First, the RCD remains zero for frequencies up to $\omega = 2\Delta_{\text{gap}}$, where $\Delta_{\text{gap}}$ is the energy gap of the spinon band.
Note, since Raman scattering is a $\bvec{q}=0$ probe, it measures the two-fermion DOS at fixed ${\bvec k}$ and $-{\bvec k}$~, which is  proportional to the rescaled single-particle DOS $\rho(2\epsilon)$. 
Therefore, the onset of the RCD is sensitive to the opening of the gap induced by TRS breaking.  
Second, the RCD signal exhibits a  jump at the onset of excitations, $ \Delta I_{\text{RCD}}$,
proportional to the discontinuity in the DOS
$\Delta \rho$ given by
\begin{align}
    \Delta I_{\text{RCD}} 
    &= \sum_{\bvec{k}_\text{D} } \big[v_\text{B}(\bvec{k}_\text{D})F_{xy}(\bvec{k}_\text{D})
    + v_\text{C}(\bm k_{\text{D}}) C(\bm k_{\text{D}}) +v_\text{S}(\bvec{k}_\text{D})\big] \Delta \rho \label{eq:jump_height} 
\end{align}
where $k_\text{D}$ denotes the momenta of the Dirac cones in the first Brillouin zone. 

\section{Application to representative chiral Spin liquid models }
\subsection{Chiral Kitaev QSL}

\begin{figure*}[t]
\includegraphics[width=1\textwidth]{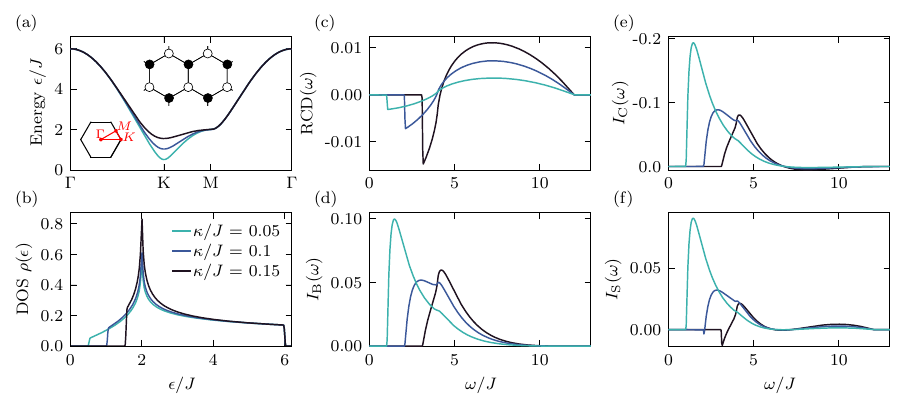}%
\caption{\label{fig:kitaev_panel}  \textbf{
RCD of the chiral Kitaev honeycomb QSL (Eq.~\eqref{eq:KHM_H}).}
(a) Matter fermion dispersion $\epsilon/J$ shown along  a high-symmetry path (red) in the Brillouin zone (black hexagon, inset).
A topological gap opens at the $\bvec{K}$ points as $\kappa$ increases.
(b) Single-particle DOS $\rho(\epsilon)$ for different values of $\kappa$.
(c) The  LMC  RCD, $I_{\text{RCD}}(\omega)$,  as function of energy transfer $\omega$.  The RCD shows a vanishing response at low energy transfer, followed by a jump at the onset, inherited from the behavior of $\rho(\epsilon)$.   $I_{\text{RCD}}(\omega)$ is the  sum of three contributions:
 $I_\text{B}(\omega)$ from the Berry curvature shown in (d), 
 $I_\text{G}(\omega)$  from  the quantum metric shown in (e),  and
  $I_\text{S}(\omega)$ from
  the second order derivatives of eigenstates shown in (f).}
\end{figure*}

Our first example is an exactly solvable chiral $\mathbb{Z}_2$ QSL, the Kitaev honeycomb model with broken TRS~\cite{Kitaev2006}
\begin{align}
\label{eq:KHM_H}
    H = -
    \sum_{\langle j ,k \rangle_\alpha } J^\alpha \sigma_j^\alpha \sigma_k^\alpha 
    - \kappa \sum_{\langle jkl \rangle_{\alpha \beta}} \sigma^{\alpha}_{j} \sigma^{\gamma}_{k} \sigma^{\beta}_{l}
\end{align}
in the topological $B$ phase with isotropic exchange $J^\alpha = 1$ for $\alpha = x,y,z$.
The first term is the bond-dependent Kitaev interaction, 
while the second term perturbatively incorporates a magnetic field that breaks TRS but preserves the exact solution~\cite{Kitaev2006}.  Bonds are denoted as $\langle jk \rangle_{\alpha}$, and $\langle jkl \rangle_{\alpha \beta}$ represents a path consisting of $\langle jk \rangle_{\alpha}$ and $\langle kl \rangle_{\beta}$ where $(\alpha\beta\gamma)$ is  a permutation of $(xyz)$.
The model  is solved by fractionalizing spins into Majorana fermions, which give rise to matter and flux excitations. 
The ground state lies in the flux-free sector \cite{Kitaev2006, Lieb1961}, 
leaving a quadratic Hamiltonian in matter fermions.

Fig.~\ref{fig:kitaev_panel}~(a) displays the matter fermion dispersion   $\epsilon(\bvec{k})$  of the Kitaev model for different values of  $\kappa$, while Fig.~\ref{fig:kitaev_panel}~(b) presents the corresponding DOS  $\rho(\epsilon)$. A finite  $\kappa$  opens a topological gap at the Dirac cones. 
The Berry curvature $F_{xy}(\bvec{k})$ peaks at  $\bvec{K}$  and  $\bvec{K}^{\prime}$ and sums up to a finite Chern number.
Similarly, different components of the quantum geometric connection $C_{\eta, \mu \nu}^{mn}(\bvec{k})$ exhibit peaks at the $\bvec{K}$-points.
Representative momentum-resolved profiles of the QGT and QGC are shown in  Fig.~\ref{fig:kitaev_qgt}, Fig.~\ref{fig:kitaev_re_qgc} and Fig.~\ref{fig:kitaev_im_qgc} in Appendix~\ref{App:qg}.
In Fig.~\ref{fig:kitaev_panel}~(c) we show the RCD response within the LMC-approach , which is finite  once TRS is broken  trough a finite $\kappa$-term.  For comparison, we show the exchange based LF approach in Fig.~\ref{fig:kitaev_lmc_vs_lf_I}. We discuss the relation between the LF and LMC Raman vertex, and the conditions under which they coincide or differ, in App.~\ref{App:LFvsLMC}.  

The features of the RCD are governed by the interplay between the density of states $\rho(\epsilon)$ and the components of the QGT. In particular, the RCD is enhanced when regions of large DOS overlap with regions of significant geometric weight in momentum space (shown in App.~\ref{App:qg}).
At higher energies, the vanishing density of states results in a cutoff of the RCD at $\omega = 12 J$.
Note that the shape of the RCD is mostly  unaffected by the strength of $\kappa$.
At low energies, the onset is shifted, but otherwise only the amplitude of the RCD, but not the shape, changes. This originates from the fact, that the $\kappa$ term is taken to be perturbatively small.
To further explore the connection between the RCD and band topology we analyze its three distinct contributions: the Berry curvature contribution  $I_\text{B}(\omega)$ in Fig.~\ref{fig:kitaev_panel}~(d), the quantum geometric connection contribution  $I_\text{C}(\omega)$ in Fig.~\ref{fig:kitaev_panel}~(e) and 
the second derivative contribution $I_\text{S}(\omega)$ in Fig.~\ref{fig:kitaev_panel}~(f). 

At low energies,  all three contributions are governed by the topology of the spinon bands. As $\kappa$ becomes nonzero, the Dirac cones gap out,  leading to a finite Berry curvature and quantum geometric connection.
For small gaps, the Berry curvature is highly localized in momentum space, resulting in large values and a prominent peak in 
the Berry contribution $I_\text{B}$
immediately after the onset
(see the cyan curve,  $\kappa = 0.05$,   in Fig.~\ref{fig:kitaev_panel}~(c)). As $\kappa$ increases, the Berry curvature spreads over the Brillouin zone, reducing its contribution to
$I_\text{B}$ (see blue curve for  $\kappa = 0.1$   and black curve for  $\kappa = 0.15$  in Fig.~\ref{fig:kitaev_panel}~(c)).
At low energies the quantum geometric connection contribution $I_\text{C}$ and  $I_\text{S}$ exhibits the same behavior as $I_\text{B}$. However, at high energies, both contributions $I_\text{C}$ and 
$I_\text{S}$  allow for a sign-change, unlike the Berry-curvature contribution $I_\text{B}$  which is strictly positive.  Note that the individual contributions $I_\text{B}$ and $I_\text{S}$ are  one order of magnitude larger than the resulting RCD.

\subsection{ Topological Trivial Kitaev QSL }\label{App:APhaseRCD}

\begin{figure*}
\includegraphics[width=1\textwidth]{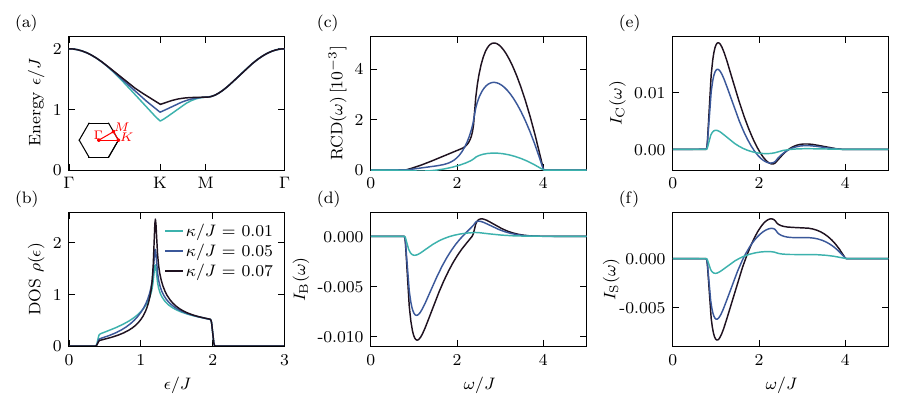}%
\caption{\label{fig:anisotropic_kitaev_panel} 
\textbf{
RCD of the topologically trivial $A$-phase of the Kitaev honeycomb QSL  at $J_x = J_y = 0.2$ and $J_z = 0.6$.}
(a) Matter fermion dispersion $\epsilon/J$ shown along  a high-symmetry path (red) in the Brillouin zone (black hexagon, inset).
(b) Single-particle DOS $\rho(\epsilon)$ for different values of $\kappa$.
(c) The RCD, $I_{\text{RCD}}(\omega)$,  as function of energy transfer $\omega$.  The RCD shows a vanishing response at low energy transfer, followed by steady onset, contrary to the  
behavior of $\rho(\epsilon)$.   $I_{\text{RCD}}(\omega)$ is the  sum of three contributions:
 $I_\text{B}(\omega)$ from the Berry curvature shown in (d), 
 $I_\text{C}(\omega)$  from  the quantum geometric connection shown in (e),  and
  $I_\text{S}(\omega)$ from
  the second order derivatives of eigenstates shown in (f). 
  }
\end{figure*}

To illustrate  that a finite RCD probes the handedness of the excitations rather than topology itself, 
we analyze the topologically trivial A phase of the Kitaev honeycomb model~\cite{Kitaev2006} with anisotropic couplings $J_x = J_y = 0.2$ and $J_z = 0.6$ and a perturbative time-reversal symmetry-breaking $\kappa$-term, see  Fig.~\ref{fig:anisotropic_kitaev_panel}. In the $A$ phase,
the system is gapped and topologically trivial, with a vanishing Chern number. Nevertheless, as shown in Fig.~\ref{fig:anisotropic_kitaev_panel}(c), the RCD response remains finite.
The nonzero RCD signal can be understood directly from its decomposition into geometric contributions. For a two-band system, the RCD intensity is given by Eq.~\eqref{eq:RCD_expansion_simple}.
In the anisotropic $A$ phase, both the Berry curvature $F_{xy}(\bm k)$ and the quantum geometric connection $C(\bm k)$ are strongly concentrated near the $\bm M$ points of the Brillouin zone. However, these geometric quantities do not enter the RCD directly. Instead, they appear multiplied by momentum-dependent prefactors, which weight their contribution differently across the Brillouin zone. 
Therefore, what controls the RCD is not the integrated Berry curvature, but its weighted distribution in momentum space. While the Berry curvature vanishes upon integration over the Brillouin zone, $\int_{\mathrm{BZ}} d^2k\, F_{xy}(\bm k) = 0$, reflecting the trivial Chern number of the $A$ phase,
the presence of the momentum-dependent weight $v_{\mathrm{B}}(\bm k)$ leads to a finite contribution
\begin{equation}
\int_{\mathrm{BZ}} d^2k\, v_{\mathrm{B}}(\bm k)\, F_{xy}(\bm k) \neq 0, 
\end{equation}
as shown in Fig.~\ref{fig:anisotropic_khm_qgt}~(b). 
 Similarly, for the quantum geometric connection 
 \begin{equation}
\int_{\mathrm{BZ}} d^2k\, v_{\mathrm{C}}(\bm k)\, C(\bm k) \neq 0,
\end{equation}
once TRS is broken,  shown in Fig.~\ref{fig:anisotropic_khm_qgt}~(c). 

Comparing the A-phase RCD Fig.~\ref{fig:anisotropic_kitaev_panel}~(c) and B-phase RCD Fig.~\ref{fig:kitaev_panel}~(c) reveals qualitatively distinct behavior that can be traced back to differences both in the density of states (DOS) and in the nature of the eigenstates. 
In the  B-phase, the opening of a topological gap at the Dirac points leads to a characteristic DOS with a sharp onset at the gap edge.   In addition, the eigenstates near the gapped Dirac cones carry strongly concentrated quantum geometric weight. As a result, the RCD exhibits a sharp onset at $\omega = 2\Delta$, with an amplitude that scales with the DOS discontinuity $\Delta \rho$, and a pronounced sign change due to competing geometric contributions.
In contrast, the trivial A-phase lacks Dirac points and exhibits a smooth DOS without a sharp feature at the gap edge. Consequently, the RCD shows a gradual onset and no sign change. 
  
Despite these differences, both phases display a finite RCD once TRS is broken. This demonstrates that the RCD probes the handedness of the excitations induced by TRS breaking rather than the spinon band topology alone. As a result, the RCD remains finite despite the vanishing Chern number, highlighting its sensitivity to the quantum geometric structure of the excitations rather than to topology alone.

\begin{figure*}
\includegraphics[width=1\textwidth]{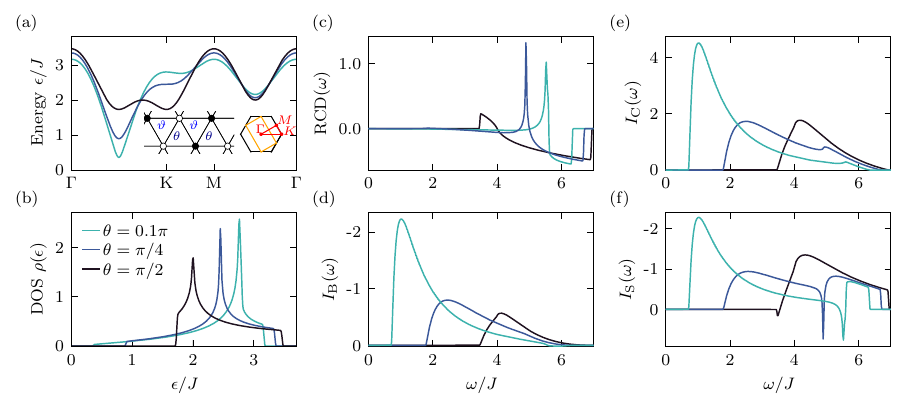}%
\caption{\label{fig:tlafh_panel} 
\textbf{RCD of the $U(1)$ chiral QSL on the triangular lattice Eq.~\eqref{eq:TLAFH_H}.} 
(a) Mean-field spinon band dispersion $\epsilon/J$ shown along a high symmetry path (red) in the BZ. The  inset shows  the original BZ (black) and 
the reduced zone due to the gauge structure (orange). 
The inset in the middle shows the flux configuration $[\vartheta, \theta] =   [\pi-\theta,\theta]$. 
(b) Single-particle density of states $\rho(\epsilon)$ for different $\theta$.
(c) The RCD as function of energy transfer $\omega$. 
The RCD shows a vanishing response at low energy transfer, followed by a jump at the onset.  
$I_{\text{RCD}}(\omega)$ is the  sum of three contributions:
$I_\text{B}(\omega)$ from the Berry curvature is shown in (d), 
$I_\text{G}(\omega)$  from  the quantum metric is shown in (e),  and
$I_\text{S}(\omega)$ from
the second order derivatives of eigenstates is shown in (f).
}
\end{figure*}

\subsection{$U(1)$ chiral QSL on the triangular lattice} 
As the second example, we consider the $J_1$-$J_2$ antiferromagnetic Heisenberg model on the triangular lattice~\cite{Hu2015,Gong2017,Iqbal2016,Wietek2017}
\begin{align}
     \label{eq:TLAFH_H}
     H  = J_1 \sum_{\langle i,j\rangle} \bvec{S}_{i} \cdot \bvec{S}_{j} + J_2 \sum_{\langle \langle i,j \rangle \rangle} \bvec{S}_i \cdot \bvec{S}_j + ...,
\end{align}
 where $J_1$  and $J_2$ are the first- and second-nearest-neighbor couplings,  and $...$ indicates additional interactions that may stabilize a chiral spin-liquid phase.  
Intensive numerical research on  this model  has shown that a chiral QSL is  stabilized in an extended region of the phase diagram,  for example by a TRS breaking scalar-chirality term $J_\chi = \sum_{i,j,k \in \Delta} \bvec{S}_i \cdot (\bvec{S}_j \times \bvec{S}_k)$,  and can be described efficiently by a simple variational wave function of partons explained in the following~\cite{Hu2015,Iqbal2016,Wietek2017,Gong2017,Zhu2018}.
We consider a standard parton
construction expressing spin operators  $\bvec{S}_i = (S_i^x, S_i^y, S_i^z) $ in terms of spinless fermionic partons
 $f_{i}$ as $
S_i^\mu = \frac{1}{2} f_{i}^\dagger \sigma^\mu_{\alpha\beta} f_{i}
$, where
$\sigma^\mu$ $(\mu = x, y, z)$ are the Pauli matrices.
The spin interaction, which is quartic in fermionic operators, can be decoupled using a suitable mean-field approximation, leading to an effective mean-field Hamiltonian
\begin{align}
    \label{eq:Dirac_QSL_MF_H}
    H_{\text{MF}} = \sum_{\langle ij \rangle} t_{ij} f_{i}^\dagger f_j + \hc,
\end{align}
where $t_{ij} = |t_{ij}|e^{i \theta_{ij}}$ denotes the hopping amplitude. 
The system breaks TRS depending on the complex phase, stabilizing a chiral QSL phase.
We choose the hopping phases $\theta$ to follow
the convention from Ref.~\cite{Song2021} such that we have a unit cell with two triangular plaquettes each carrying flux  $[\vartheta, \theta] = [\pi-\theta,\theta]$
as shown in the inset of Fig.~\ref{fig:tlafh_panel}. 
For $\theta=0$, the system realizes the staggered $\pi$-flux phase, featuring a fermionic dispersion with two Dirac points, characteristic of the well-known Dirac QSL.
Figure~\ref{fig:tlafh_panel}~(a) shows the spinon dispersion $\epsilon(\bvec k)$ of the upper band for different flux configurations,  while the corresponding single-particle DOS is shown in Fig.~\ref{fig:tlafh_panel}~(b).
Tuning away from the $[\pi,0]$ Dirac QSL, any $\theta \neq 0$ breaks TRS and a topological gap opens, driving the Dirac $U(1)$ QSL into a gapped chiral QSL characterized by two Chern bands of spinons with $C=\pm1$. 
The components of the quantum geometric tensor reach their largest absolute values near the gapped Dirac cones and along the paths connecting them, details are shown in  Fig.~\ref{fig:tlafh_qgt}.

Fig.~\ref{fig:tlafh_panel}~(c) shows the full RCD response which is only nonzero for finite $\theta$, as expected. 
First, we  connect features of the RCD to features in the DOS. Again, due to particle-hole symmetry and energy conservation, features in the single-particle DOS at energy $\epsilon$ need to be identified with features at $\omega/2$ in the RCD. It includes the vanishing RCD from the topological gap and its onset aligning with jumps/peaks in the spinon DOS.
Second, the RCD shape depends on  $\theta$, as it influences not only the band dispersion but also the matrix elements and QGT Eq.~\eqref{eq:RCD_expansion_simple}.
The positions of the Van Hove singularity and the zero crossing in the RCD shift with 
$\theta$. This strong dependence is different from the Kitaev chiral QSL, where different values of $\kappa$ only modulate the amplitude of the RCD, but do not change the shape.  
Third, the main features of the RCD can again be traced back to the contributions of $I_\text{B}, I_\text{C}$ and $I_\text{S}$.
For small gaps, the Berry curvature and QGC  have large values  around the gapped Dirac cones, similar to the Kitaev QSL leading to large contributions to $I_\text{B}$ and $I_\text{C}$, see Fig.~\ref{fig:tlafh_panel}~(d) and (e). The interplay of the energetic spread of the Berry curvature and the location of the Van Hove singularity leads to sign changes of the RCD. Interestingly, because regions of larger DOS overlap with large Berry curvature and quantum metric contributions, the full RCD of the chiral $U(1)$ QSL on the triangular lattice is two orders of magnitude larger compared to the Kitaev case. 
In both cases, we showed how the interplay of the spinon quantum geometric quantities governs the LMC Raman response.  

\section{Discussion and Outlook} 
We computed the RCD response for two QSLs with broken TRS and finite scalar spin 
chirality -- the Kitaev honeycomb model in a magnetic field and a triangular lattice QSL.  For the effective quasiparticle descriptions considered in this work, we identified the conditions under which the LF and LMC formalisms yield equivalent Raman vertices.
In particular, the two approaches yield equivalent Raman vertices when the effective 
spinon kinetic energy is generated by the same exchange interactions that define the 
LF Raman operator. Establishing this equivalence allows us to transfer the direct 
connection between the Raman response and quasiparticle quantum geometry obtained 
within the LMC framework to the corresponding LF Raman response.
Additional contributions to the quasiparticle Hamiltonian originating from microscopic processes beyond the standard exchange process,  such as the field-induced $\kappa$ term in the Kitaev model, require separate consideration.

Using the LMC framework, we decompose the RCD into three contributions associated with the Berry curvature, the quantum geometric connection, and higher-order momentum derivatives. These contributions dominate the low-energy response, underscoring the central role of spinon quantum geometry. A key result is that, the spinons exhibit quantum geometric effects in the Raman response analogous to electronic systems, despite being charge-neutral excitations in Mott insulators. Since only the total RCD is experimentally accessible, it is desirable to develop protocols that disentangle the underlying geometric contributions.

Our results also clarify the interpretation of the RCD:  while a finite signal requires broken TRS, it is not a direct measure of topological invariants such as the Chern number. Instead, the RCD reflects the handedness encoded in the quantum geometry of the excitations, and can therefore remain finite even in topologically trivial phases.

With this, our work lays the foundation for a systematic understanding of the RCD  in systems with fractionalized quasiparticles,  
opening up several directions for future research. Incorporating finite-$T$ effects, in particular vison excitations~\cite{Nasu2016}, as well as disorder and the extension of quantum geometry beyond translationally invariant systems are natural next steps. Going beyond the free-spinon approximation in both the triangular lattice and perturbed Kitaev models is essential, especially in view of recent RCD experiments in $\alpha$-RuCl$_3$~\cite{sahasrabudhe2024chiral}.
Beyond QSL, our method can be transferred to fractionalized phases with intrinsic angular momentum, such as fractional Chern insulators. 

\vspace{0.2cm}
Numerical data and simulation codes are available on the online repository Zenodo \cite{zenodo} upon reasonable request.

\vspace{0.5cm}
\section{ Acknowledgment} 
We thank Emil Boström, Silvia Viola Kusminskiy and Jonas Habel for helpful discussions. 
E.K. acknowledges support from the Institute for Advanced Study (IAS) of the Technical University of Munich and hospitality of the University of Minnesota, Minneapolis. 
V.L. acknowledges support from the ``Studienstiftung des Deutschen Volkes''.
N.B.P. was supported by the U.S. Department   of Energy, Office
 of Science, Basic Energy Sciences under Award No. DE-SC0018056. N.B.P. also
acknowledges  partial support of the Alexander von Humboldt Foundation.
J.K.  acknowledges support from the Deutsche Forschungsgemeinschaft (DFG, German Research Foundation) under Germany’s Excellence Strategy–EXC–2111–390814868, TRR 360 – 492547816 and DFG grants No. KN1254/1-2 and No. KN1254/2-1, as well as the Munich Quantum Valley, which is supported by the Bavarian state government with funds from the Hightech Agenda Bayern Plus. J.K.  acknowledges support from the TUM-Imperial flagship partnership. J.K.  and N.B.P. thank the hospitality of Aspen Center for Physics, which is supported by National Science Foundation grant PHY-2210452.

\twocolumngrid

\appendix
\section{Analytical expressions for the RCD contributions}\label{App:ExpansionRCD}
\renewcommand{\theequation}{A\arabic{equation}}
\renewcommand{\thefigure}{A\arabic{figure}}
\renewcommand{\thetable}{A\arabic{table}}
\setcounter{equation}{0}
\setcounter{figure}{0}
\setcounter{table}{0}

Here, we derive Eq.~\eqref{eq:RCD_in_LMC} and present the analytical expressions for  different contributions to the RCD in Eq.~\eqref{eq:RCD_expansion_simple}.
The Raman intensity is given by the Raman correlator as $I(\omega)= \int dt e^{i\omega t}\langle \mathcal{T}_t R^\dagger(t)R(0)\rangle$ with $\langle . \rangle$ the ground state average. To calculate the RCD, we use the equivalence between the LF and LMC Raman operators (see App.~\ref{App:LFvsLMC}). For a generic fermionic multi-level system, we use the matrix notation for the LMC Raman operator in Eq.~\eqref{eq:def_LMC_R} 
\begin{align}
    \label{eq:def_LMC_R_operator}
    L^{(2)}_{\mu \nu}(\bm k,t)  =  \sum_{m,n}\psi^\dagger_m(\bvec{k},t) 
    l_{mn,\mu\nu}^{(2)}(\bvec{k})
    \psi_n(\bvec{k},t),
\end{align}
with $l_{mn,\mu\nu}^{(2)}(\bvec{k}) = P_m(\bm k) \big[ \partial_{\mu} \partial_{\nu}h(\bvec{k})\big] P_n(\bm k)$.  For non-interacting systems, the time dependence follows from the Heisenberg picture.

Contracting over the polarization vector  indices $\mu,\nu = x,y$ and using the hermiticity in the operator indices,the symmetry $L^{(2)}_{\mu\nu}=L^{(2)}_{\nu\mu}$, and
the time-independence of $l^{(2)}_{mn,\mu \nu}(\bm k)$,  we obtain the RCD in terms of LMC operators:
\begin{small}
\begin{align}
\label{eq:multiband_RCD}
I_{\mathrm{RCD}}(\omega)
=
&\int \mathrm{d}t\, e^{i\omega t}
\left[
\langle [R^{+-}_{\mathrm{LMC}}(t)]^\dagger R^{+-}_{\mathrm{LMC}}(0) \rangle
-
\langle [R^{-+}_{\mathrm{LMC}}(t)]^\dagger R^{-+}_{\mathrm{LMC}}(0) \rangle
\right] \nonumber \\
=
-2 &\int \mathrm{d}t\, e^{i\omega t}\,
\mathrm{Im}
\left\langle
\sum_{\bm k,\bm q}
\big(L^{(2)}_{xx}(\bm k,t)-L^{(2)}_{yy}(\bm k,t)\big)
L^{(2)}_{xy}(\bm q,0)
\right\rangle. 
\end{align}
\end{small}
We  next transform to the band-diagonal basis using Eq.~\eqref{eq:def_LMC_R_operator}. For a generic multiband system, this yields
\begin{widetext}
\begin{equation}
    I_\text{RCD}(\omega) =   2 \sum_{\bm k}  \text{Im} \text{Tr} 
     \left[
    \left( P_m(\bm k)  l_{xx}^{(2)}(\bm k)   P_n(\bm k)  - P_m (\bm k) l_{yy}^{(2)} (\bm k) P_n(\bm k) \right) P_n(\bm k) l_{xy}^{(2)}(\bm k) P_m(\bm k) \right] \left[n_f(\epsilon_m(\bm k)) \right]\Big[1 - n_f(\epsilon_n(\bm k))\Big] \delta(\omega - \epsilon_{nm}(\bm k)  ),
    \label{eq:RCD_projector}
\end{equation}
\end{widetext}
where the trace runs over band indices, $n_f$ is the Fermi function, and $\epsilon_{nm}(\bm k) =\epsilon_n(\bm k) -\epsilon_m(\bm k) $.

To relate Eq.~\ref{eq:RCD_projector} to band geometry, we rewrite the LMC matrix elements in terms of derivatives acting on the projectors. In the eigenstate representation, this relation is derived using the Hellmann-Feynman theorem, as discussed in in Ref~\cite{Topp2021}. Ref~\cite{Mitscherling2025} discusses the corresponding relations using the multi-band projector formalism.

We now specialize to a two-band system, as realized in the spinon mean-field Hamiltonians considered here. At zero temperature, the Fermi factors  restrict the sum to interband transitions $m\neq n$  and selects  $m=1$ the lower and $n=2$ the upper band.  For a non-degenerate  two band system, the identity
\begin{align}
\label{eq:two_band_pe}
P_m (\partial_\mu \partial_\nu h ) P_n
&= \epsilon_{nm} P_m (\partial_\mu \partial_\nu P_n) P_n 
+(\partial_\nu \epsilon_{nm}) P_m (\partial_\mu P_n) P_n \nonumber \\
&\quad+ (\partial_\mu \epsilon_{nm}) P_m (\partial_\nu P_n) P_n
\end{align}
allows us to express the RCD entirely in terms of derivatives of the band projectors.
 Substituting Eq.~(\ref{eq:two_band_pe}) into the two-band limit of Eq.~\eqref{eq:RCD_projector}, we obtain 
\begin{widetext}
\vspace{-10pt}
\begin{align} 
I^\text{RCD}(\omega)  &\sim 2 \sum_\bm{k} \text{Im} \text{Tr} \Big[\enm P_n (\px \py P_n) P_m  +\px \enm P_n (\py P_n) P_m   + \py \enm P_n (\px P_n) P_m  \Big] 
\Big[ \epsilon_{nm} P_m (\px \px P_n) P_n + 2 \px \enm P_m (\px P_n) P_n \nonumber \\ 
 &- \epsilon_{nm} P_m (\py \py P_n) P_n + 2 \py \enm P_m (\py P_n) P_n \Big] \delta(\omega - \enm) 
\end{align} 
\vspace{-10pt}
\end{widetext}
modulo the temperature factors selecting the band indices. 
The product results in 12 terms. For concreteness, consider one of the terms 
\begin{align}
\label{eq:connection_to_simplify}
\text{Im} \text{Tr} \Big[ \enm P_m (\px \px P_n) \px \enm P_n (\py P_n) P_m\Big]
\end{align}
In the two-band limit, using $\mathds{1}=P_n+P_m$, its derivatives, $(\partial_y P_n)P_m = P_n(\partial_y P_n)$, and cyclicity of the trace, Eq.~\eqref{eq:connection_to_simplify} reduces to
$\epsilon_{nm} (\partial_x \epsilon_{nm})\, \text{Im}\,\mathrm{Tr}\Big[ P_n (\partial_y P_n) (\partial_x \partial_x P_n) \Big]
= \epsilon_{nm} (\partial_x \epsilon_{nm})\, \text{Im}\, C^{nn}_{yxx}$. 
Evaluating all twelve terms, we find that two vanish by projector orthogonality, two correspond to the Berry curvature, six to the quantum geometric connection, and two to higher-order derivative terms
 \begin{widetext}\vspace{-10pt}
\begin{align}
I^{\mathrm{RCD}}(\omega)
&= 2 \sum_{\bm k}
\bigg[
\underbrace{\left[ (\partial_x \epsilon_{nm})^2 + (\partial_y \epsilon_{nm})^2 \right] F^{nm}_{xy}}_{\text{Berry curvature}}\,-\,
\underbrace{2(\partial_x \epsilon_{nm}) \epsilon_{nm}\, \mathrm{Im}\, C^{nm}_{xxy}
+ 2(\partial_y \epsilon_{nm}) \epsilon_{nm}\, \mathrm{Im}\, C^{nm}_{xyy}}_{\text{QGC}}
\nonumber\\
&\quad
\underbrace{+ (\partial_x \epsilon_{nm}) \epsilon_{nm}\, \mathrm{Im}\, C^{nn}_{yxx}
+ (\partial_y \epsilon_{nm}) \epsilon_{nm}\, \mathrm{Im}\, C^{nn}_{xxx}
- (\partial_x \epsilon_{nm}) \epsilon_{nm}\, \mathrm{Im}\, C^{nn}_{yyy}
- (\partial_y \epsilon_{nm}) \epsilon_{nm}\, \mathrm{Im}\, C^{nn}_{xyy}}_{\text{QGC}}
\nonumber\\
&\quad
\underbrace{+ \epsilon_{nm}^2\, \mathrm{Im}\,\mathrm{Tr}\!\left[P_m (\partial_x \partial_x P_n) P_n (\partial_x \partial_y P_n)\right]
- \epsilon_{nm}^2\, \mathrm{Im}\,\mathrm{Tr}\!\left[P_m (\partial_y \partial_y P_n) P_n (\partial_x \partial_y P_n)\right]}_{\text{higher derivatives}}
\bigg]
\delta(\omega - \epsilon_{nm}) .
\end{align}
\vspace{-10pt}
\end{widetext}
 We group these terms into three contributions,
\begin{equation}
\mathcal{M}_{\mathrm{RCD}}(\bm k)
= \nu_B(\bm k)\, F_{xy}(\bm k)
+ \nu_C(\bm k)\, C(\bm k)
+ \nu_S(\bm k),
\end{equation}
where the prefactors $\nu_B$, $\nu_C$, and $\nu_S$ encode the dependence on the band energies and their derivatives, while $F_{xy}$ and $C$ denote the Berry curvature and quantum geometric connection contributions, respectively.
We show the three different contributions to the RCD as well as their sum momentum resolved for the Kitaev model and in for the chiral $U(1)$ QSL as well as the components of the QGT and QGC in Appendix~\ref{App:qg}.

\section{Scope of the LF--LMC correspondence } 
\label{App:LFvsLMC}

\renewcommand{\theequation}{B\arabic{equation}}
\renewcommand{\thefigure}{B\arabic{figure}}
\renewcommand{\thetable}{B\arabic{table}}
\setcounter{equation}{0}
\setcounter{figure}{0}
\setcounter{table}{0}

Here we  discuss the conditions under which  the LMC and LF Raman vertices are equivalent. 
 The correspondence holds when the effective spinon
Hamiltonian is generated by the same microscopic exchange processes that
also determine the LF Raman vertex.  Under this condition, the two
constructions yield the same Raman vertex, up to an overall sign. The correspondence becomes incomplete when the effective quasiparticle
Hamiltonian contains additional terms that do not originate from the
exchange processes entering the LF construction. In such cases,
\begin{align}
h(\bm k)=h_{\rm exch}(\bm k)+h_{\rm other}(\bm k), 
\end{align}
and the LMC and LF vertices generally differ, 
\begin{align} R^{ss^\prime}_{\rm LMC} = -R^{ss^\prime}_{\rm exch} -R^{ss^\prime}_{\rm other}, \qquad R^{ss^\prime}_{\rm LF} = R^{ss^\prime}_{\rm exch}. 
\end{align} 

We note that a recent work~\cite{Liu2026} develops a related LMC approach for magnons and discusses the LMC-LF equivalence. 

\subsection{Kitaev Honeycomb model}

The magnetic field-induced $\kappa$ term provides an instructive example of the scope of the LF--LMC correspondence. In the effective Majorana description, the $\kappa$ term contributes directly to the spinon kinetic Hamiltonian and therefore enters the LMC Raman vertex through the momentum dependence of $h(\bm k)$. Its microscopic origin, however, is a Zeeman coupling rather than an exchange interaction generated by virtual electron hopping processes.

We illustrate this by first considering the exchange-generated component of the Kitaev Hamiltonian. In this case, the LF Raman vertex follows directly from the exchange interactions entering Eq.~\eqref{eq:KHM_H}
\renewcommand{\arraystretch}{1.3}
\begin{small}
\begin{align}
    \label{eq:R_LF_complex_fermion_basis}
    R^{ss^\prime}_\text{LF} = \sum_{\mu\nu}  \epsilon^{s}_\mu  \epsilon^{*s^\prime}_\nu \sum_{\bm k}  
    \begin{pmatrix} f^\dagger_{\bm k}  & f_{- \bm k} \end{pmatrix}
    \begin{pmatrix} 
    r_{11}^{\mu\nu}(\bm k) & - r_{12}^{\mu\nu}(\bm k ) \\
        -r_{12}^{*\mu\nu}(\bm k ) & - r_{11}^{\mu\nu}(\bm k ) 
    \end{pmatrix}
    \begin{pmatrix} f_{\bm k}   \\ f^\dagger_{- \bm k} \end{pmatrix}
\end{align}
\end{small}
\renewcommand{\arraystretch}{1.0}
with $r_{11}^{\mu \nu}(\bm k ) = \text{Re} \sum_{\substack{d_x,d_y,d_z} } J^{\alpha}d_\mu^i d_\nu^i e^{i \bm k  \cdot \bm d_\alpha }$ and $r_{12}^{\mu \nu}(\bm k ) =  - i \text{Im} \sum_{\substack{d_x,d_y,d_z} } J^{\alpha}d_\mu^i d_\nu^i e^{i \bm k  \cdot \bm d_\alpha }$.

Although the Zeeman coupling does not generate a Raman vertex within the standard exchange-based LF construction, a magnetic field dependent Raman vertex may arise from higher-order microscopic processes. Since such processes are not considered by the standard LF formalism, their form and prefactor are not fixed by the coefficient $\kappa$ entering the effective Majorana Hamiltonian. To consider this possibility, we introduce an additional Raman contribution parametrized by $\kappa'$ as 
\begin{small}
\begin{align}
    R^{ss^\prime}_{\text{other}} = 
    \sum_{\mu\nu}  \epsilon^{s}_\mu  \epsilon^{*s^\prime}_\nu \sum_{\bm k}  
    \begin{pmatrix} f^\dagger_{\bm k}  & f_{- \bm k} \end{pmatrix}
    \begin{pmatrix} 
    0 & - r_{\text{other},12}^{\mu\nu}(\bm k ) \\
        -r_{\text{other}, 12}^{*\mu\nu}(\bm k ) & - 0 
    \end{pmatrix}
    \begin{pmatrix} f_{\bm k}   \\ f^\dagger_{- \bm k} \end{pmatrix}
\end{align}
\end{small}
with 
$r_{\text{other}, 12}^{\mu \nu}(\bm k ) = 
2 \kappa' \sum_{\substack{d_1,d_3,d_5} }d_\mu^i d_\nu^i \sin(\bm k \cdot \bm d_i)$.
In general, the phenomenological parameter $\kappa'$ is independent of the parameter $\kappa$ entering the effective Majorana Hamiltonian. 
The special cases $\kappa'=0$ and $\kappa'=\kappa$ correspond to the standard exchange-based LF construction and the extended LF construction reproducing the LMC result.

For the LMC Raman operator we consider the flux-free sector of the Kitaev honeycomb model with the $\kappa$ term, Eq.~\eqref{eq:KHM_H}. After Fourier transformation and introducing complex fermions it takes the form \cite{Kitaev2006}
\begin{align}
\label{eq:kitaev_hamiltonian}
    H = \sum_{\bm k}  \psi^\dagger_{\bm k}  h(\bm k) \psi_{\bm k} = \sum_{\bm k}
    \begin{pmatrix} f^\dagger_{\bm k} & f_{-\bm k}\end{pmatrix}
    \begin{pmatrix} \xi_{\bm k} & - \Delta_{\bm k}\\
        -\Delta_{\bm k}^* & - \xi_{\bm k}
    \end{pmatrix}
    \begin{pmatrix} f_{\bm k}  \\ f^\dagger_{-\bm k}\end{pmatrix}.
\end{align}
Here, $\xi_{\bm k}=\mathrm{Re}\,S(\bm k)$ and $\Delta_{\bm k}=-T(\bm k)-i\,\mathrm{Im}\,S(\bm k)$, where
$S(\bm k)=\sum_{\alpha=x,y,z} J^\alpha e^{i\bm k\cdot\bm d_\alpha}$ and
$T(\bm k)=-2\kappa\sum_{i=1,3,5}\sin(\bm k\cdot\bm d_i)$.
The nearest-neighbor vectors are
$\bm d_x=\tfrac{1}{2}(\sqrt{3},1)^T$,
$\bm d_y=\tfrac{1}{2}(-\sqrt{3},1)^T$,
$\bm d_z=(0,-1)^T$, and the next-nearest-neighbor vectors are
$\bm d_1=(\sqrt{3},0)^T$,
$\bm d_2=\tfrac{1}{2}(\sqrt{3},3)^T$,
$\bm d_3=\tfrac{1}{2}(-\sqrt{3},3)^T$.

The LMC Raman vertex finally follows from the second derivatives of the Hamiltonian,
$l^{(2)}_{\mu\nu}(\bm k)=\partial_{\mu}\partial_{\nu} h(\bm k)$.
Evaluating these derivatives for the Hamiltonian in Eq.~\eqref{eq:kitaev_hamiltonian} reproduces the exchange and other contribution introduced above 
\begin{align}  
R^{ss^\prime}_{\text{LMC}} = - R^{ss^\prime}_{\rm exch} - R^{ss^\prime}_{\rm{other}},
\end{align}
which coincides with the extended LF construction upon  identifying $\kappa =\kappa'$.
\begin{figure}[h!]
\includegraphics[width=0.49\textwidth]{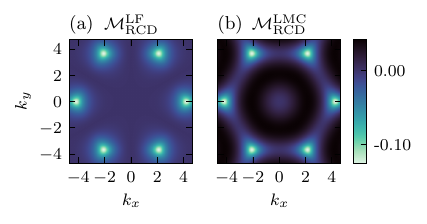}%
\caption{\label{fig:kitaev_lmc_vs_lf_M}   
\textbf{Comparison of RCD matrix elements in the isotropic Kitaev model.}
(a) Standard exchange-based LF matrix elements.
(b) LMC matrix elements obtained from the effective Majorana Hamiltonian. 
Both approaches exhibit similar weight near the gapped Dirac cones at $\mathbf K$ and $\mathbf K'$, where the dominant quantum-geometric contributions are concentrated. Differences arise away from the Dirac points due to the field-induced $\kappa$ contribution contained in the LMC description but absent from the standard LF construction.}
\end{figure}

\begin{figure*}[t]
\includegraphics[width=\textwidth]{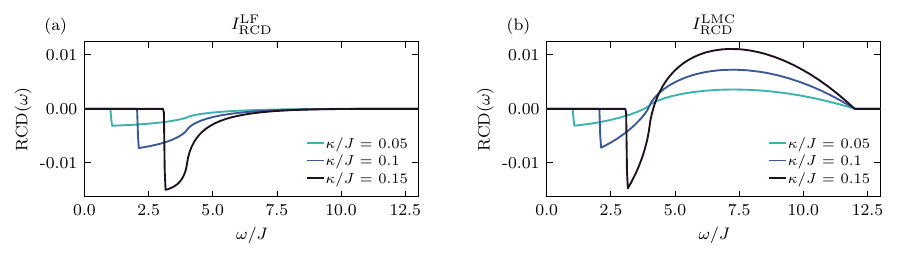}%
\caption{\label{fig:kitaev_lmc_vs_lf_I}   \textbf{Comparison of RCD spectra in the isotropic Kitaev model.} (a) RCD calculated from  exchange-based LF Raman vertex ($\kappa'=0$). (b) RCD obtained from LMC Raman vertex. At low energies, both responses are dominated by excitations near the gapped Dirac cones and therefore show similar behavior. At intermediate and high energies, differences emerge as the LMC Raman vertex includes the field-induced contribution associated with the $\kappa$ term, whereas the exchange-generated LF Raman vertex does not, illustrating the limitations of the LF--LMC correspondence when the effective quasiparticle Hamiltonian contains contributions that are not generated by exchange interactions.   }
\end{figure*}

This demonstrates that the exchange-generated component of the Kitaev Hamiltonian satisfies the LF--LMC correspondence discussed above. The field-induced $\kappa$ contribution, however, enters the effective Majorana Hamiltonian through a distinct microscopic mechanism and therefore lies outside the scope of the standard LF construction. The comparison in Fig.~\ref{fig:kitaev_lmc_vs_lf_M} and Fig.\ref{fig:kitaev_lmc_vs_lf_I} illustrates how this additional contribution modifies the Raman matrix elements and the resulting RCD spectrum.   

\subsection{Triangular lattice chiral QSL}
For the triangular-lattice model, we focus on the dominant $J_1$-$J_2$ exchange contributions to derive the Raman vertex.
Using the hopping phases in Ref.~\cite{Song2021},
the momentum space Hamiltonian reads  
\begin{align}
\label{eq:tlafh_h_ab_basis}
    H = \sum_k \begin{pmatrix} a^\dagger_{\bm k} & b^\dagger_{\bm k} \end{pmatrix} 
    \begin{pmatrix}
        h_{11}(\bm k ) & h_{12}(\bm k ) \\ h_{21}(\bm k ) & - h_{11}(\bm k ) 
    \end{pmatrix}
    \begin{pmatrix} a_{\bm k} \\ b_{\bm k}  \end{pmatrix} ,
\end{align}
with $h_{11}(\bm k) = -2\cos(\theta -\bm k  \cdot \bm d_y)$, $h_{21}(\bm k) =h^*_{12}\bm k)$  and  $h_{12}(\bm k) = -2\cos(\theta - \bm k \cdot \bm d_x)$  $- 2i \cos\left( \bm k\cdot (\bm d_x + \bm d_y) + \theta\right)$.
The nearest neighbor lattice vectors are 
$ \bm d_x= (1,  0)^T$ and $\bm d_y  = \frac{1}{2}(-1, \sqrt{3})^T$.

 The LF Raman vertex in Eq.~\eqref{eq:def_LF_R} takes the form
\renewcommand{\arraystretch}{1.3}
\begin{align}
\label{eq:TLAFH_LF_R}
 R^{ss^\prime}_\text{LF} = \sum_{\mu\nu}  \epsilon^{s}_\mu  \epsilon^{*s^\prime}_\nu \sum_{\bm k}  \begin{pmatrix} a^\dagger_{\bm k} & b^\dagger_{\bm k} \end{pmatrix} 
    \begin{pmatrix}
        r_{11}^{\mu \mu}(\bm k) & r_{12}^{\mu \nu}(\bm k) \\ r_{12}^{* \mu \nu}(\bm k)) & - r_{11}^{\mu \nu}(\bm k) 
    \end{pmatrix}
    \begin{pmatrix} a_{\bm k} \\ b_{\bm k} \end{pmatrix} 
\end{align}
\renewcommand{\arraystretch}{1.0}
with 
$r_{11}^{\mu \nu} (\bm k) = -\cos(\theta - \bm k \cdot \bm d_y ) d_y^\mu d_y^\nu$
and 
$r_{12}^{\mu \nu} (\bm k) = -\cos(\theta - \bm k \cdot \bm d_x )d_x^\mu d_x^\nu \nonumber   + i \cos( \bm k\cdot (\bm d_x+\bm d_y) + \theta) (d_x+d_y)^\mu (d_x+d_y)^\nu $

The LMC Raman vertex follows from the second derivatives of the Hamiltonian,
$l^{(2)}_{\mu\nu}(\bm k)=\partial_{k_\mu}\partial_{k_\nu} h(\bm k)$.
Evaluating these derivatives for Eq.~\eqref{eq:tlafh_h_ab_basis} reproduces Eq.~\eqref{eq:TLAFH_LF_R}, up to an overall sign.

\section{Symmetry constraints} 
\label{App:symcon}
\renewcommand{\theequation}{C\arabic{equation}}
\renewcommand{\thefigure}{C\arabic{figure}}
\renewcommand{\thetable}{C\arabic{table}}
\setcounter{equation}{0}
\setcounter{figure}{0}
\setcounter{table}{0}

In the following we show that the second-derivative contributions are generically nonzero under $\bm k$ summation once TRS is broken, i.e. they are symmetry allowed. We rewrite the second-derivative expression as 
\begin{align}
   \label{eq:supp_is}
    I_{\text{S}} (\omega ) &\sim   \sum_{\bm k}  \Im \text{Tr} \left[
    P_1 \left( \left( 
    \partial_{k_x}^2- \partial_{k_y}^2\right) P_2 \right) P_2 \left( \partial_{k_x}\partial_{k_y}  P_2 \right)   \right].
\end{align}
Since momentum transforms the same as position under point group operations and derivatives transform as the corresponding coordinates, the relevant transformation property is described by the polynomial $(k_x^2 - k_y^2)k_xk_y$. 

Note that the momentum dependence of the RCD integrand transforms as the polynomial
$(k_x^2-k_y^2)k_xk_y$, while the integrand of the Berry curvature contribution transforms as $(k_x^2+k_y^2)k_xk_y$. The same symmetry reasoning applies to the terms involving
$C^{mn}_{\eta;\mu\nu}$. This object transforms as a rank-three
momentum-space tensor, with the same symmetry as $ \sim k_{\eta} k_{\mu} k_{\nu}$ under point group operations. 

\subsection{Isotropic Kitaev honeycomb model.}
\label{subsec:kitaev_symmetry}
For the point group $D_{3d}$ of the ideal Kitaev honeycomb model we choose the set of generators 
\begin{align} 
C_3^z = \begin{pmatrix} 
-\frac{1}{2} & - \frac{\sqrt{3}}{2} & 0 \\
\frac{\sqrt{3}}{2}& - \frac{1}{2}& 0 \\ 
0 & 0 & 1
\end{pmatrix} 
&&
C_2^y = \begin{pmatrix}
    -1 & 0 & 0 \\ 
    0 & 1 & 0 \\ 
    0 & 0 & -1 
\end{pmatrix}
&& 
i =  \begin{pmatrix}
    -1 & 0& 0\\ 
    0& -1 &0 \\
    0& 0 & -1 
\end{pmatrix}
\nonumber 
\end{align} 
The inversion acts trivially on the in-plane polynomial  $(k_x^2 - k_y^2)k_xk_y$, since it contains an even number of momentum factors. The rotation  $C_2^y$ maps $(k_x^2 - k_y^2)k_xk_y \to  -(k_x^2 - k_y^2)k_xk_y$, such that the matrix elements of $I_\text{S}$ sum to zero, in agreement with the constraint from TRS.

When TRS is broken, the ordinary point group becomes a 
magnetic point group, in which the $C^y_2$ 
rotation is no longer a symmetry by itself,
but remains a symmetry when combined with TRS: $C^y_2 \to C^y_2  \mathcal{T}$. 
Under $\mathcal{T}$, the momentum reverses $\bm k \to -\bm k$, which leaves the even-order momentum factors unchanged.  However, the complex conjugation accompanying  TRS flips the sign of the imaginary part of the  matrix elements in Eq.~\eqref{eq:supp_is}, thereby canceling the minus sign generated by the $C_2^y$ rotation. As a result, the combined symmetry $C_2^y  \mathcal{T}$ no longer enforces a vanishing Brillouin zone sum. 

If the $C_3$ rotation remains a symmetry of the magnetic point group, it only mixes
$(k_x^2 - k_y^2)k_xk_y$ with its partner functions within the two-dimensional irrep.
It does not map the
Brillouin-zone sum to its negative and therefore does not require the sum to
vanish.
The remaining point-group symmetries, obtained by combining the generators, do not introduce any further cancellations.  

Therefore, once TRS is broken, the magnetic point group does not constrain the value of $I_s$ in Eq.~\eqref{eq:supp_is} to zero.

\subsection{Chiral U(1) QSL on the triangular lattice} 
The point group of the $[\pi,0]$ flux pattern  is $D_{3}$. We choose the following generators 
\begin{align} 
C_3^z = \begin{pmatrix} 
-\frac{1}{2} & - \frac{\sqrt{3}}{2} & 0 \\
\frac{\sqrt{3}}{2}& - \frac{1}{2}& 0 \\ 
0 & 0 & 1
\end{pmatrix} 
&&
C_2^y = \begin{pmatrix}
    -1 & 0 & 0 \\ 
    0 & 1 & 0 \\ 
    0 & 0 & -1 
\end{pmatrix}
\nonumber 
\end{align}
As in the Kitaev case, the twofold rotation $C_2^y$ maps $(k_x^2 - k_y^2)k_xk_y \to - (k_x^2 - k_y^2)k_xk_y$, such that the Brillouin zone sum vanishes, consistent with the TRS constraint. 
In addition, since $C_2^y$ changes $z \to -z$, it also reverses the orientation of the flux $\phi$, mapping $\phi$ to $-\phi$. In the $[\pi,0]$ sector this maps to $[-\pi,0]$, which is equivalent to the original flux pattern, as the flux is only defined mod $2\pi$. 

In the chiral flux sector $[\pi-\theta,\theta]$ with $\theta \neq 0$, $C_2^y$ alone maps $[\pi-\theta,\theta] \to [-\pi+\theta,-\theta]$, which is not equivalent to the original flux pattern. However, the combined  antiunitary symmetry operator $C_2^y \mathcal{T}$ preserves the flux pattern, since 
\begin{align} 
C_2^y \mathcal{T}[\pi-\theta,\theta] \to C_2^y[-\pi+\theta,-\theta] \to   [\pi-\theta,\theta].
\end{align} 
At the same time, the complex conjugation accompanying $\mathcal{T}$
changes the sign of the imaginary part of the matrix elements in
Eq.~\eqref{eq:supp_is}.
 This  additional minus sign cancels the minus sign generated by the $C_2^y$ rotation. As a result, the combined symmetry $C_2^y  \mathcal{T}$ no longer enforces the Brillouin zone sum to vanish, as it maps the flux pattern and the polynomial back to itself.  Consideration for the $C_3$ rotation and the other symmetry operations are analogous to Section~\ref{subsec:kitaev_symmetry}. 

\section{Quantum geometric quantities and matrix elements} 
\label{App:qg}
\renewcommand{\theequation}{D\arabic{equation}}
\renewcommand{\thefigure}{D\arabic{figure}}
\renewcommand{\thetable}{D\arabic{table}}
\setcounter{equation}{0}
\setcounter{figure}{0}
\setcounter{table}{0}

In the main text, we showed that the RCD decomposes into three contributions: $v_\text{B}(\bvec{k})F_{xy}(\bvec{k})$, linked to the Berry curvature, $v_\text{C}(\bvec{k})C(\bm k)$, related to the quantum geometric connection and $v_\text{S}(\bvec{k})$, associated with second order derivative objects. We show the matrix elements and the underlying geometric objects, namely the components of the quantum geometric tensor~\cite{Berry1984,Provost1980}
\begin{align}
\label{eq:sup_QGT}
Q^{mn}_{\mu \nu }(\bvec{k}) =\text{tr} \left[P_n(\bm k) \partial_\mu P_m(\bm k) \partial_{\nu} P_n(\bm k)  \right] = g^{mn}_{\mu \nu}(\bvec{k}) - \frac{i}{2} F^{mn}_{\mu \nu}(\bvec{k})
\end{align}
and the real and imaginary part of the quantum geometric connection~\cite{ahn2022riemannian,Mitscherling2025}
\begin{align}
\label{eq:sup_QGC}
C^{mn}_{\eta; \mu \nu}(\bm k)  = 
\Tr\left[
P_n(\bm k) \partial_{\mu}P_m(\bm k) 
\left( \partial_{\eta} \partial_{\nu} P_n (\bm k)+ \partial_{\eta} P_m(\bm k) \partial_{\nu}P_n(\bm k)  \right) 
\right],
\end{align}
for both models.

For a two-band system $\mathds{1}_2 = P_1 + P_2$ with non-degenerate bands, the band-resolved components of the QGT Eq.~\eqref{eq:sup_QGT}  are not independent. They satisfy~\cite{Mitscherling2025}  
\begin{align} 
g^{12}_{\mu \nu}(\bm k) = - g^{22}_{\mu\nu}(\bm k) =  - g^{11}_{\mu\nu}(\bm k) =  g^{21}_{\mu\nu}(\bm k), \\ 
F^{12}_{\mu \nu}(\bm k) = - F^{22}_{\mu\nu}(\bm k) =  F^{11}_{\mu\nu}(\bm k) =  -F^{21}_{\mu\nu}(\bm k)
\end{align} 
and for the quantum geometric connection~\cite{Mitscherling2025}
\begin{align}
C^{12}_{\eta; \mu \nu}(\bm k) = - C^{22}_{\mu; \eta \nu}(\bm k) = \left( - C^{11}_{\mu; \eta \nu}(\bm k) \right)^* = \left( C^{21}_{\eta; \mu \nu} (\bm k)  \right)^*.
\end{align}
A single independent component determines all band-resolved quantities. As in the main text,
$P_1$ projects onto the lower band and $P_2$ projects onto the upper band, and we will display $Q_{\mu \nu}^{21}(\bm k)$ and $C_{\eta; \mu \nu}^{21}(\bm k)$ for both models.

\subsection{Isotropic Kitaev honeycomb model.}  
\begin{figure*}
    \includegraphics[width=1\linewidth]{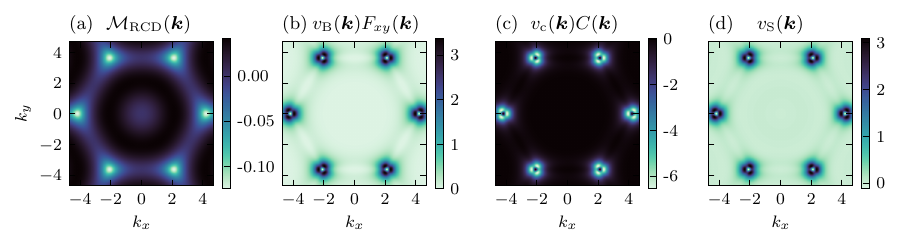}%
    \caption{\label{fig:khm_matrix_elements}
    \textbf{Contributions to the RCD of the chiral Kitaev honeycomb QSL at $\kappa /J = 0.05$.}
     The full RCD matrix elements $\mathcal{M}_\text{RCD}(\bvec{k})$ are shown in (a), which are the sum of the matrix elements $v_\text{B}(\bvec{k})F_{xy}(\bvec{k})$ shown in~(b), $v_\text{C}(\bvec{k})C(\bm k)$ shown in~(c) and $v_\text{S}(\bvec{k})$ shown in~(d). The contributions in~(b) and~(c) show features at the $\bvec{K}$-point, originating from the Berry curvature and quantum geometric connection. }
\end{figure*}

\begin{figure*}
    \centering
    \includegraphics[width=1\linewidth]{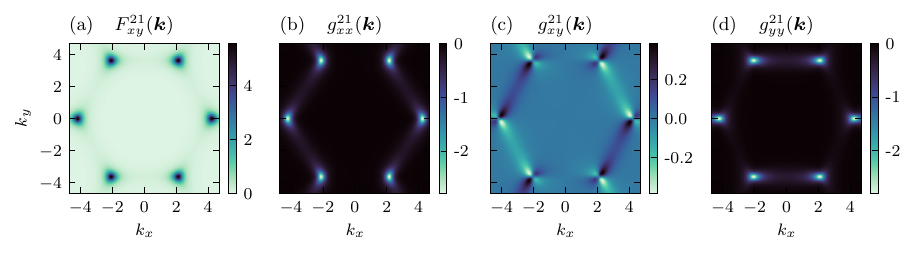}
    \caption{\textbf{QGT of the chiral Kitaev honeycomb QSL at $\bm \kappa/J = 0.05$}. (a)-(d) show the components of the QGT, with the Berry curvature in (a) and the components of the quantum metric in (b)-(d). Significant changes in the QGT occur at the $\bm K$-points, where the Dirac cones gap out topologically.}
    \label{fig:kitaev_qgt}
\end{figure*}

\begin{figure*}
    \centering
    \includegraphics[width=1\linewidth]{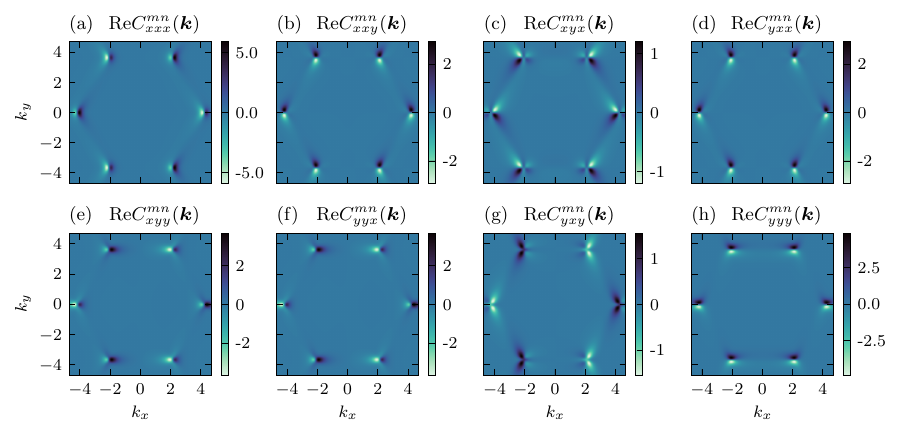}
    \caption{\textbf{Real part of the QGC of the chiral Kitaev honeycomb QSL at $\bm \kappa/J = 0.05$}. (a)-(h) show the different components of the QGC. Significant changes in the QGC occur at the $\bm K$-points, where the Dirac cones gap out topologically.}
    \label{fig:kitaev_re_qgc}
\end{figure*}

\begin{figure*}
    \centering
    \includegraphics[width=1\linewidth]{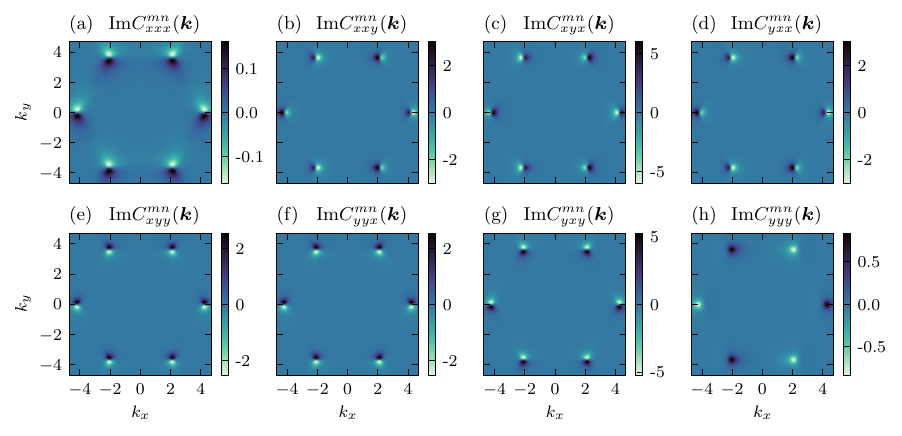}
    \caption{\textbf{Imaginary part of the QGC of the chiral Kitaev honeycomb QSL at $\bm \kappa/J = 0.05$}. (a)-(h) show the different components of the QGC. Significant changes in the QGC occur at the $\bm K$-points, where the Dirac cones gap out topologically.}
    \label{fig:kitaev_im_qgc}
\end{figure*}

Fig.~\ref{fig:khm_matrix_elements} shows the matrix elements of the RCD (panel~(a)) and the different contributions (panel~(b)-(d)), while Fig.~\ref{fig:kitaev_qgt} shows the components of the quantum geometric tensor, both for the isotropic Kitaev honeycomb model at a field of $\bm \kappa/J = 0.05$.  The Berry curvature and quantum metric are strongly concentrated at the $\bm K$ points, where the perturbative $\kappa$-term gaps out the Dirac cones and generates a finite Chern number. With increasing gap, the quantum geometry spreads into the BZ. The QGC components, shown in Fig.~\ref{fig:kitaev_re_qgc} (real part) and Fig.~\ref{fig:kitaev_im_qgc} (imaginary part), are similarly peaked at the $\bm K$-points, reflecting the spatial variation of the quantum geometry near the topological gap. 
The matrix elements Fig.~\ref{fig:khm_matrix_elements}~(b) and (c) inherit the features from the underlying Berry curvature and connection, especially visible at the peaked values around the $\bm K$-points.


\begin{figure*}[h]
\includegraphics[width=1\linewidth]{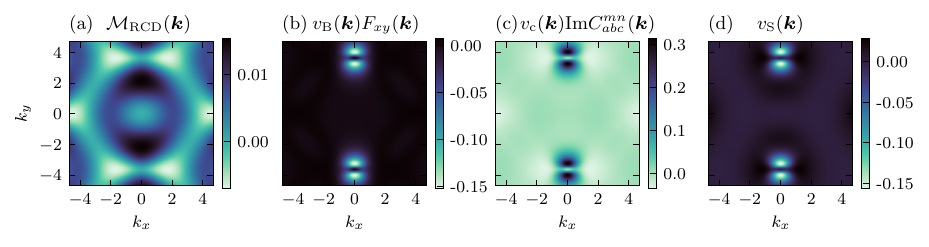}%
\caption{\label{fig:anisotropic_khm_qgt}
\textbf{Contributions to the RCD of the A-phase of the Kitaev honeycomb QSL at $\kappa /J = 0.05$ and $J_x = J_y = 0.2$, $J_z = 0.6$}
The full RCD matrix elements $\mathcal{M}_\text{RCD}(\bvec{k})$ are shown in (a), which are the sum of the matrix elements $v_\text{B}(\bvec{k})F_{xy}(\bvec{k})$ shown in~(b), $v_\text{C}(\bm k) C(\bm k)$ shown in ~(c) and $h_\text{S}(\bvec{k})$ shown in~(d).
} 
\end{figure*}

\subsection{Chiral U(1) QSL on the triangular lattice.}
\begin{figure*}
    \includegraphics[width=1\linewidth]{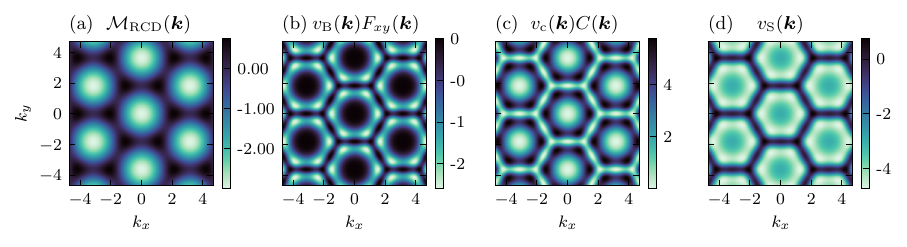}%
    \caption{\label{fig:tlafh_matrix_elements}
    \textbf{Contributions to the RCD of the chiral $U(1)$ QSL on the triangular lattice for the flux-pattern $[\pi/2,\pi/2]$.}
    The full RCD matrix elements $\mathcal{M}_\text{RCD}(\bvec{k})$ are shown in (a), which are the sum of the matrix elements $v_\text{B}(\bvec{k})F_{xy}(\bvec{k})$ shown in~(b), $v_\text{C}(\bm k) C(\bm k)$ shown in~(c) and $v_\text{S}(\bvec{k})$ shown in~(d). 
     }
\end{figure*}

\begin{figure*}
    \centering
    \includegraphics[width=1\linewidth]{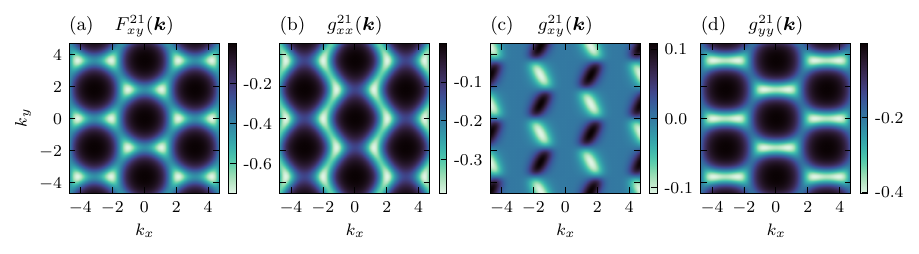}
    \caption{\textbf{QGT of the chiral $U(1)$ QSL on the triangular lattice for the flux pattern $[\pi/2, \pi/2]$.} (a)-(d) show the components of the QGT, with the Berry curvature in (a) and the components of the quantum metric in (b)-(d). The main change in the QGT occurs, where the Dirac cones gap out topologically, as well as along the lines connecting the $\bm K$-points. }
    \label{fig:tlafh_qgt}
\end{figure*}

\begin{figure*}
    \centering
    \includegraphics[width=1\linewidth]{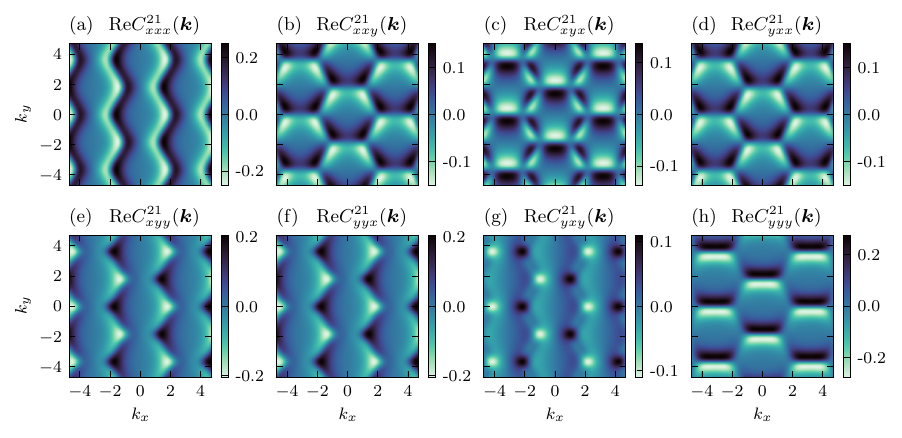}
    \caption{\textbf{Real part of the QGC of the chiral $U(1)$ QSL on the triangular lattice for the flux pattern $[\pi/2, \pi/2]$.} (a)-(h) show the components of the QGC.  The main change in the QGC occurs, where the Dirac cones gap out topological, as well as along the lines connecting the $\bm K$-points.}
    \label{fig:tlafh_re_qgc}
\end{figure*}

\begin{figure*}
    \centering
    \includegraphics[width=1\linewidth]{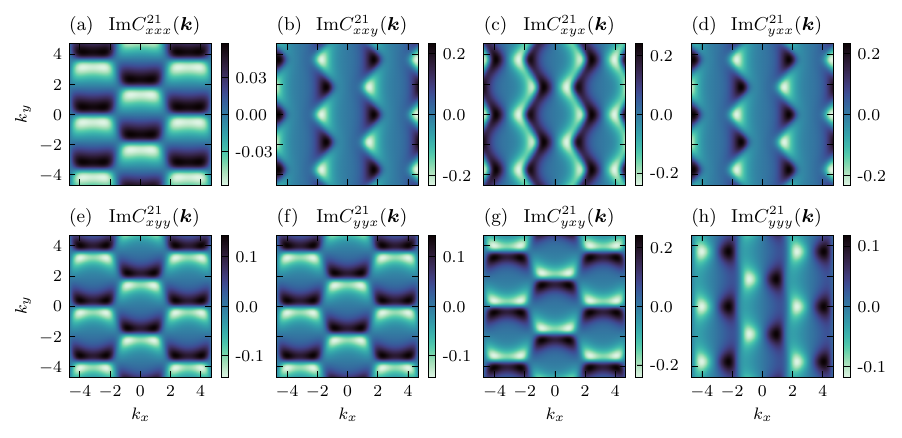}
    \caption{\textbf{Imaginary part of the QGC of the chiral $U(1)$ QSL on the triangular lattice for the flux pattern $[\pi/2, \pi/2]$.} (a)-(h) show the components of the QGC.  The main change in the QGC occurs, where the Dirac cones gap out topologically, as well as along the lines connecting the $\bm K$-points.}
    \label{fig:tlafh_im_qgc}
\end{figure*}

Fig.~\ref{fig:tlafh_matrix_elements} shows the matrix elements of the RCD (panel~(a)) and the different contributions (panel~(b)-(d)),
while Fig.~\ref{fig:tlafh_qgt} shows the components of the quantum geometric tensor, both for the flux configuration $[\pi/2, \pi/2]$. As in the Kitaev case, the main features are concentrated where the Dirac cones gap out, but the quantum geometric weight extends along the paths connecting the gapped Dirac cones in the BZ. This extended distribution, visible also in the QGC (Fig.~\ref{fig:tlafh_re_qgc} and~\ref{fig:tlafh_im_qgc}), leads to a larger overlap between the quantum geometry and connection and the regions of high density of states, visible in the enhanced matrix elements Fig.~\ref{fig:tlafh_matrix_elements}. This is the origin of the enhancement of the RCD signal compared to the Kitaev case.


\section{Numerical details}
\label{App:num}
\renewcommand{\theequation}{E\arabic{equation}}
\renewcommand{\thefigure}{E\arabic{figure}}
\renewcommand{\thetable}{E\arabic{table}}
\setcounter{equation}{0}
\setcounter{figure}{0}
\setcounter{table}{0}

Numerical evaluation of matrix elements and $\bm k$ sums uses a $800\times800$ $\bm k$-mesh. The delta distribution is approximated by
\begin{align}
\delta_\epsilon(x) = \frac{1}{\sqrt{2 \pi \epsilon}}e^{\frac{-x^2}{2 \epsilon} }.
\end{align}

For the isotropic Kitaev honeycomb model, we discretize the energy in $400$ slices over $\Omega \in [0, 6.2J]$ for the DOS and over  $\Omega \in [0,13J]$ for the intensities. We  calculated the intensities with the delta-sequence broadening  $\epsilon_K= 1.8\times 10^{-4}$

For the anisotropic Kitaev honeycomb model,  we
discretize the energy in $400$ slices over  $\Omega \in [0, 3J]$ for the DOS and over  $\Omega \in [0,5J]$ for the intensities. We  calculate the intensities with the delta-sequence broadening  $\epsilon_K= 1.8\times 10^{-4}$

For the $U(1)$ chiral QSL on the triangular lattice we discretise the energy in $400$ slices over  $\Omega \in [0, 3.7J]$ for the DOS and over $\Omega \in [0,7J]$ for the intensities. We  calculated the intensities with the delta-sequence broadening $\epsilon_K= 1.34\times 10^{-4}$.

The code is available upon reasonable request under~\cite{zenodo}.

\bibliography{main}

\end{document}